# Spin transition in $LaCoO_3$ investigated by resonant soft X-ray emission spectroscopy

M. Magnuson, S. M. Butorin, C. Såthe and J. Nordgren

*Department of Physics, Uppsala University, P. O. Box 530,*

*S-751 21 Uppsala, Sweden*

P. Ravindran

*Department of Chemistry, University of Oslo, P. O. Box 1033 Blindern,*
*N-0315 Oslo, Norway*

**Abstract**

The spin transition in $LaCoO_3$ is investigated by temperature-dependent resonant soft X-ray emission spectroscopy near the Co $2p$ absorption edges. This element-specific technique is more bulk sensitive with respect to the temperature induced spin-state of the $Co^{3+}$ ions in $LaCoO_3$ than other high-energy spectroscopic methods. The spin transition is interpreted and discussed with *ab-initio* density-functional theory within the fixed-spin moment method, which is found to yield consistent spectral functions to the experimental data. The spectral changes for $LaCoO_3$ as a function of temperature suggest a change in spin-state as the temperature is raised from 85 to 300 K while the system remains in the same spin state as the temperature is further increased to 510 K.

## 1  Introduction

The interest in transition metal oxides has grown significantly in several branches of solid-state physics. Recent investigations have shown that some of these systems exhibit a variety of fascinating properties e.g., high-temperature superconductivity in the cuprates and colossal magneto-resistance in the manganates [1]. For the cobaltates, one of the most important phenomena is the temperature-induced spin-transition in $LaCoO_3$ [2,3,4]. This material is unique in the sense that it undergoes one or more spin-state transition(s) with increasing temperature and has therefore attracted attention as a compound with peculiar magnetic properties. Although the history of the issue of the spin-transition(s) in $LaCoO_3$ is quite long, the controversy has not yet completely settled down. One reason is difficulties of precise electronic structure measurements at high-temperatures above $\sim 500$ K. Depending on the method used, the literature reveals several conflicting interpretations with respect to which temperature the spin state transition(s) takes place. Studies using valence-band X-ray photoelectron spectroscopy (XPS) [6,5] and X-ray absorption spectroscopy (XAS) [7] have been performed at different temperatures up to about $\sim 600$ K. In a quasi-ionic picture, the ground state of $LaCoO_3$ at 4.2 K has $Co^{3+}$ ions in the $3d^6$ configuration which form a singlet S=0, $t_{2g}^6$ low-spin (LS) state. The changes in the spectra have been interpreted to be due to thermal population into a S=2, $t_{2g}^4 e_g^2$ high-spin (HS) state from the LS state [8,7,5]. However, both XPS and XAS are surface sensitive methods, which may influence the spectral structures. In addition, the observed spectral changes through the expected magnetic transition at ∼ 90 K are much smaller than those predicted by calculations. The experiments show significant changes in the spectra only at much higher temperatures between 400-650 K which is difficult to reconcile with bulk-sensitive magnetic susceptibility measurements [11,12,13,9,10].

On the other hand, calculations within the local density approximation LDA+U [14] have shown that the HS metallic state appears at much higher temperatures than 600 K. Therefore, it has been proposed that the spin-state transition(s) in LaCoO$_3$ occurs in two different steps involving an S=1, $t_{2g}^5 e_g^1$ intermediate-spin (IS) state [15,14]. However, although LDA calculations are generally *ab initio*, LDA+U calculations are not fully *ab initio* since a parameter is used. Moreover, simulations of 80 K and 300 K XPS valence band spectra using parametric multiplet calculations have shown that relatively small spectral changes can be expected between these temperatures [16]. However, the XPS valence-band experiments in Ref [16] only dealt with a temperature of 80 K and the technique is known to be surface sensitive. The nature of the spin-transition(s) is therefore not yet settled and needs further investigation. With the combination of both bulk-sensitive electronic-structure measurements up to not less than 500 K and fully *ab initio* band-structure calculations, one can expect to obtain reliable information about the nature of the spin-state transition(s) of LaCoO$_3$.

In this Letter we investigate the electronic structure and the temperature dependence of LaCoO$_3$ using resonant soft X-ray emission (RSXE) spectroscopy with selective excitation energies around the Co 2$p$ thresholds. This technique is more bulk sensitive than other spectroscopic techniques and each atomic element is separately probed by tuning the excitation energy to the appropriate core edge. The RSXE spectroscopy follows the dipole selection rule and conserves the charge-neutrality of the probed system. By varying the temperature at the appropriate choice of incident photon energies, it is demonstrated that the RSXE technique is sensitive to spin-state transitions when the density-of-states (DOS) differ. The experimental results are interpreted and discussed with the combined analysis of accurate state-of-the-art full-potential density-functional calculations within the generalized gradient correction and the fixed-spin moment method. Recently, electronic structure properties of perovskites were reliably predicted with this *ab-initio* method [17]. The combination of the bulk-sensitive RSXE spectroscopy and the band-structure calculations presented here should give reliable insight into the issue of the spin-transition(s) in LaCoO$_3$.

## 2 Experimental

The measurements were performed at beamline BW3 at HASYLAB, Hamburg, using a modified SX700 monochromator [18]. The Co $L_{2,3}$ RSXE spectra were recorded using a high-resolution grazing-incidence grating spectrometer with a two-dimensional position-sensitive detector [19]. During the RSXE measurements at the Co 2$p$ edges, the resolution of the beamline monochromator was about 0.5 eV. The RSXE spectra were recorded with a spectrometer resolution better than ∼ 0.5 eV. All the measurements were performed with a base pressure lower than $5 \times 10^{-9}$ Torr. In order to minimize self-absorption effects [20], the angle of incidence was about 25$^o$ during the emission measurements. The emitted photons were always recorded at an angle, perpendicular to the direction of the incident photons, with the polarization vector parallel to the horizontal scattering plane to minimize the elastic contribution. Cooling to ∼ 85 K was achieved by liquid nitrogen and warming to ∼ 510 K by using an electrical coil heater on the sample holder. The sample was made of well-characterized sintered pellets [21].

# 3 *Ab initio* calculation of soft x-ray emission spectra

The electronic structure calculations were performed *ab initio* based on the spin-polarized, density-functional theory (DFT). We have applied the full-potential linearized-augmented plane wave (FPLAPW) method as embodied in the *WIEN97* code [22] using a scalar relativistic version without spin-orbit coupling. The effects of exchange and correlation are treated within the generalized-gradient-corrected local spin-density approximation using the parameterization scheme of Perdew *et al* [23]. The unit cell of $LaCoO_3$ has a rhombohedrically distorted pseudo cubic perovskite structure consisting of 2 formula units (f.u.) each containing 1 La, 1 Co and 3 O atoms. To ensure convergence for the Brillouin zone integration, 110 **k**-points in the irreducible wedge of the first Brillouin zone of the lattice were used with lattice parameters corresponding to 4.2 K. The different spin states were obtained using the fixed-spin-moment method [24,25,26] calculating the total energy as a function of constrained magnetic moments. The magnetic moments of the $Co^{3+}$ ions are 0, 2 and 4 μB/f.u. for the LS, IS and HS spin-states, respectively. The calculated RSXE spectra of $LaCoO_3$ in the different spin-states were made within the dipole approximation from the FPLAPW [22] partial DOS along the description of Neckel *et al.* [27] with the same core-to-valence matrix elements for the different spin-states.

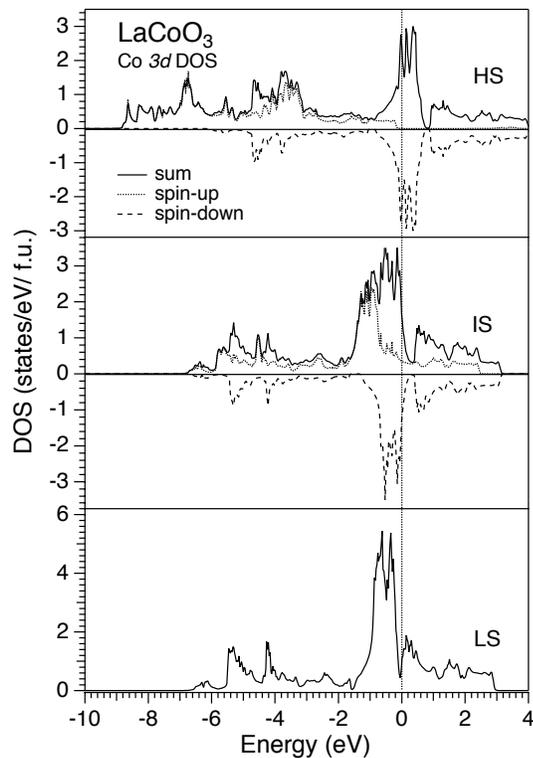

**Figure 1:** Calculated spin-dependent Co 3*d* density-of-states (DOS) for the HS (high-spin), IS (intermediate-spin) and LS (low-spin) states in $LaCoO_3$.

The core-hole and final state life-times and the instrumental resolution were simulated using a Lorentzian with a full-width-half-maximum (FWHM) of 0.32 eV, a variable Lorentzian broadening increasing from the Fermi level ($E_F$) and a 0.5 eV FWHM Gaussian, respectively. More details about fixed-spin moment calculations can be found elsewhere [28].

## 4 Results

Figure 1 shows our calculated spin-dependent DOS (states/eV/f.u.) for the constrained spin-moments of the HS (top), IS (middle) and LS (bottom) $3d$ states of $Co^{3+}$ in $LaCoO_3$. In the nonmagnetic LS ground state, the spin up and spin-down density-of-states (DOS) are identical. In the LS state, the $E_F$ is lying in a deep valley identified as a pseudo gap feature, probably indicating a semi-metallic DOS [16] and hence a semiconducting behavior. For the spin-polarized IS and HS states, the spin-up and spin-down DOS differ significantly from each other. The spin-down states have considerably more weight close to the $E_F$ than the spin-up states. As observed, the pseudo gap near the $E_F$ does not exist in the IS and HS states. In the IS state, the $E_F$ is located on the slope of the spin-down $3d$ band and the sum of the spin-bands is broader but less intense than in the case of the LS state. In the HS state, the $E_F$ is located on a peak of the spin down $3d$ DOS while the weight of the spin-up states are close to zero.

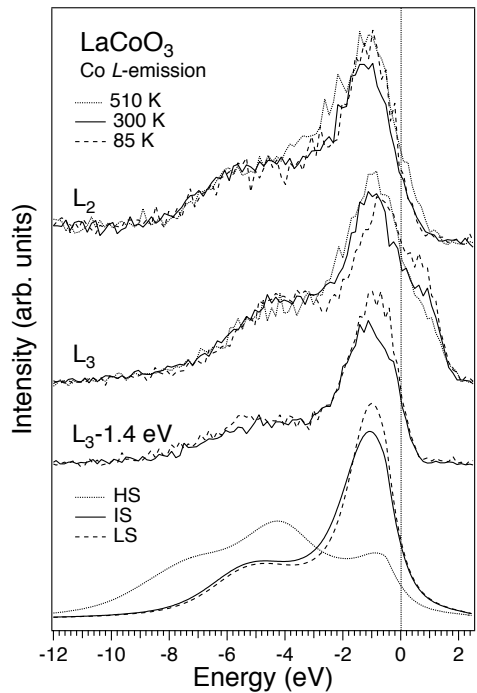

**Figure 2:** Comparison between measured and calculated RSXE spectra of Co in $LaCoO_3$. The experimental data were obtained at 85 K, 300 K and 510 K. The energy scales at each temperature were shifted to the $E_F$ using the $L_3$ and $L_2$ core-level binding energies, respectively. The calculated spectra are the $3d$ Co DOS projected with the $2p-3d$ core-to-valence dipole matrix-elements for the LS (low spin), IS (intermediate spin) and HS (high spin) states.

Figure 2 shows the results of our temperature-dependent RSXE measurements at the Co $L_2$ and $L_3$ thresholds in comparison with the calculated spectra which also include the $2p-3d$ matrix elements and the Co DOS shown in Fig. 1. The energies of the incident photon beam were set to 1.4 eV below the $L_3$ absorption maximum, and at the $L_3$ and $L_2$ maxima. The measured spectra are normalized to the incoming photon flux and plotted on a common energy scale with respect to the $E_F$ using $2p_{3/2}$ (779.5 eV) and $2p_{1/2}$ (794.5 eV) core-level photoemission binding energies [29]. As observed, the Co $L_{2,3}$ final states in the RSXE spectra of LaCoO$_3$ appear rather delocalized which makes band-structure calculations suitable for the interpretation of the spectra. The observed spectral features are thus mainly due to ordinary (normal) emission while elastic scattering only influence the spectra for the excitation energy at the $L_3$ absorption maximum. In more localized systems, where elastic scattering is more prominent, resonant inelastic X-ray scattering (RIXS) features may be identified as tracking the excitation energy. This is not observed for the delocalized Co states in LaCoO$_3$. In order to keep the discussion and the issue of the temperature dependent spin-transition in LaCoO$_3$ simple, we have chosen not to include and discuss spectra above the thresholds, where satellite structures appear.

## 5 Discussion

As observed by the experiment, the spectral shapes significantly depend on the temperature for the three different excitation energies. In the temperature-dependent RSXE spectra excited at the $L_3$ absorption maximum, the spectral shape also depends on the contribution of elastic scattering i.e., when the emitted photons have the same energy as the incoming photons. For 3$d$ metal systems in general, the elastic scattering contribution resonates and is generally stronger at the $L_3$ threshold than at the $L_2$ threshold. In Fig. 2, the elastic scattering contribution in the spectra excited at the $L_3$ absorption maximum is observed as a shoulder on the high-energy side of the main peak although the major part of this contribution has been effectively suppressed in our experimental geometry. As the temperature is raised from 85 K (dashed curve) to 300 K (full curve) at the $L_3$ absorption maximum, the main peak is shifted towards lower energy and its intensity at the high-energy shoulder of the spectrum decreases. The difference in the elastic scattering contribution as a function of temperature at the $L_3$ absorption resonance implies a redistribution of the total spectral weight towards the main peak for the spectrum measured at 300 K in comparison to the 85 K spectrum. This implies that if the elastic contribution was absent in the 85 K spectrum, the intensity of the main peak would be higher than at 300 K. This is clearly observed for the excitation energy at 1.4 eV below the $L_3$ absorption maximum and at the $L_2$ absorption maximum. At this excitation energy, the contribution of the elastic scattering can be totally neglected at all temperatures and the observation of the difference in intensity of the main peak as a function of temperature is facilitated. Taking the elastic contribution into account, the intensity of the main peak is thus significantly lower at 300 K than at 85 K for all three excitation energies.

A comparison between the calculated LS, IS and HS spectra at the bottom, and the observed experimental spectral changes, strongly suggest a change of spin-state of LaCoO$_3$ as the temperature is raised from 85 to 300 K. The spectral changes between 85 K and 300 K show that the relative amount of Co$^{3+}$ ions in the LS-state decreases and Co$^{3+}$ ions in a different spin-state are populated. The calculated HS spectrum has a completely different shape and a much lower intensity at the $E_F$ in comparison to the LS spectrum while the IS spectrum is more similar in line shape to the LS spectrum. However, the intensity of the IS spectrum is significantly lower than the LS spectrum and there is a small energy shift towards lower energy. As a consequence, it can be anticipated that if the valence-band can be described as a superposition of LS and HS spin states, the spectral shapes would show completely different spectral changes with temperature as shown by parametric multiplet models [16]. However, although the intensities are significantly lower and the spectral structures are slightly shifted to lower energies at 300 K compared to 85 K, the general profile of the spectra remains. This is consistent with the relatively small difference in line shape between the calculated LS and IS spectra ruling out the HS state.

The changes of the RSXE spectral structures as a function of temperature thus show a significant change in the spin-state of $LaCoO_3$ between 85 and 300 K which supports previous assumptions based on a LS-IS rather than a LS-HS type of spin transition below 300 K. This is in contrast to previous interpretations using parametric models of the more surface sensitive x-ray absorption and photoemission methods[7, 5]. In $LaCoO_3$, the energy of the IS state relative to the ground state is largely affected by the hybridization between the Co-$3d$ levels and the O-$2p$ band. Previous not fully *ab initio* band structure calculations within the LDA+U approach show a crossover between the LS-IS states at lattice parameters corresponding to a temperature of ~ 150 K while the IS-HS crossover occurs at much higher temperatures, above 600 K [14]. Our self-consistent *ab initio* full-potential fixed-spin-moment DFT calculations indicate that there exist a metastable IS state for temperatures above the LS-IS crossover. The onset of the LS-IS crossover appears around 290 K (i.e. 25 meV/f.u.). The calculated total energy difference between the LS state and the local minimum corresponding to the metastable IS state is 32 meV/f.u. suggesting that $LaCoO_3$ completely transforms from the LS to the IS state around 371 K.

As the temperature is raised to 510 K in the experimental RSXE spectra (dotted lines in Fig. 2), the changes in line-shape are less dramatic than the spectral changes between the 85 K and the 300 K spectra. However, at 510 K, the intensity of the main peak is significantly higher and it is slightly broader than at 300 K but there is no significant energy shift. At the $L_3$ absorption maximum, the intensity of the elastic contribution is further but less lowered compared to the 300 K spectrum. The spectral change in this temperature region is completely different from the calculated HS state, indicating that $LaCoO_3$ remains in the IS state at 510 K. According to our full-potential *ab-initio* band structure calculations, the total energy of the HS spin-state is always much higher than the IS state at all temperatures. The crossover between the IS and HS states appears for lattice parameters corresponding to a temperature of 13043 K. For lattice parameters corresponding to 4.2 K, the difference in total energy between the LS and HS states is as high as 1113 meV/f.u. (12916 K). This indicates that the probability of a population of the HS state is very low. Considering both the experimental RSXE spectral shape modifications as a function of temperature as well as the calculated total energies and temperatures at the LS-IS and IS-HS crossovers, we conclude that the system remains in the IS state as the temperature is raised to 510 K. We find that the temperature induced physical properties of $LaCoO_3$ are determined by the difference in the DOS of the LS and IS states near the $E_F$ in combination with the small total energy difference between these states. Our experimental findings combined with the theoretical results thus implies that there is no spin-transition around 500 K in $LaCoO_3$. However, the spectral change with a more intense and a somewhat broader $3d$ band indicates that a semiconductor-to-metal transition or a localized-to-itinerant electron transition takes place in this temperature region as observed by the calculations. The interpretation of the LS-IS spin-transition presented here is supported by recent parametric modelling of magnetic susceptibility measurements[30]. The complementarities between bulk-sensitive experimental results and *ab initio* calculations will be generally useful for further investigations of magnetic systems and can be considered as being more reliable than studies based solely on parametric models.

## 6 Conclusions

In summary, we have shown that the bulk-sensitive resonant soft X-ray emission spectroscopy technique is sensitive for detecting spin-states when differences can be detected in the density-of-states. As an example, we investigated the transition metal perovskite $LaCoO_3$ which exhibits a spin-state transition consistent with *ab initio* band structure calculations within the fixed spin-moment method. The agreement of the combined study between the experimental and theoretical spectra show a change of spin-state below 300 K from low-spin to intermediate-spin. At 510 K, the system remains in the intermediate-spin state ruling out the high-spin state. The utilization of specific excitation energies and resonant conditions in resonant X-ray emission experiments implies that detailed spin-state information may be obtained rather generally in various systems important for modern nano technology and spintronics.

## 7 Acknowledgements

This work was supported by the Swedish Research Council and the Göran Gustafsson Foundation for Research in Natural Sciences and Medicine.

## References

[1] M. Imada, A. Fujimori and Y. Tokura; Rev. Mod. Phys. **70**, 1039 (1998) and references therein.
[2] J. B. Goodenough, *Progress in Solid State Chemistry*, Vol. 5, p 145, Ed. by H. Reiss (Pergamon Press Inc., New York, 1967);
[3] J. B. Goodenough, *Magnetism and the Chemical Bond*, (Interscience and Wiley, New York, 1963).
[4] P. M. Raccah and J. B. Goodenough, Phys. Rev. **155**, 932 (1967); J. Appl. Phys. **39**, 1209 (1968).
[5] G. Thornton, I. W. Own and G. P. Diakun; J. Phys. Condens. Matter **3**, 417 (1991).
[6] S. R. Barman and D. D. Sarma; Phys. Rev. B **49**, 13979 (1994).
[7] M. Abbate, J. C. Fuggle, A. Fujimori, L. H. Tjeng, C. T. Chen, R. Potze, G. A. Sawatzky, H. Eisaki and S. Uchida; Phys. Rev. B **47**, 16124 (1993).
[8] S. Masuda, M. Aoki, Y. Harada, H. Hirohashi, Y. Watanabe, Y. Sakisaka and H. Kato; Phys. Rev. Lett. **71**, 4214 (1993).
[9] K. Asai, O. Yokokura, M. Suzuki, T. Naka, T. Matsumoto, H. Takahashi, N. Mori and K. Kohn; J. Phys. Soc. Jpn; **66**, 967 (1997).
[10] K. Asai, A. Yoneda, O. Yokokura, J. M. Tranquada, G. Shirane, K. Kohn; J. Phys. Soc. Jpn. **67**, 290 (1998).
[11] S. Yamaguchi, Y. Okimoto, H. Taniguchi and Y. Tokura; Phys. Rev. B **53**, R2926 (1996).
[12] M. Itoh, M. Sugahara, I. Natori and K. Motoya; J. Phys. Soc. Jpn. **64**, 3967 (1995).
[13] K. Asai, P. Gehring, H. Chou and G. Shirane; Phys. Rev. B **40**, 10982 (1989).
[14] M. A. Korotin, S. Yu. Ezhov, I. V. Solovyev, V. I. Anisimov, D. I. Khomskii and G. A. Sawatzky; Phys. Rev. B **54**, 5309 (1996).

[15] M. A. Senaris-Rodriguez and J. B. Goodenough, J. Solid State Chem. **116**, 224 (1995).
[16] T. Saitoh, T. Mizokawa, A. Fujimori, M. Abbate, Y. Takeda, M. Takano; Phys. Rev. B **55**, 4257 (1997).
[17] P. Ravindran, A. Kjekshus, H. Fjellvåg, A. Delin, O. Eriksson; Phys. Rev. B **65**, 064445 (2002).
[18] T. Möller; Synchrotron Radiat. News. **6**, 16 (1993).
[19] J. Nordgren and R. Nyholm; Nucl. Instr. Methods **A246**, 242 (1986); J. Nordgren, G. Bray, S. Cramm, R. Nyholm, J.-E. Rubensson, and N. Wassdahl; Rev. Sci. Instrum. **60**, 1690 (1989).
[20] S. Eisebitt, T. Böske, J.-E. Rubensson and W. Eberhardt; Phys. Rev. B **47**, 14103 (1993).
[21] A. N. Petrov, O. F. Kononchuk, A. V. Andreev, V. A. Cherepanov, P. Kofstad; Solid State Ionics **80**, 189 (1995); A. N. Petrov, V. A. Cherepanov, O. F. Kononchuk, L. Ya. Gavrilova; J. Solid State Chem. **87**, 69 (1990).
[22] P. Blaha, K. Schwarz and J. Luitz, *WIEN97, A Full Potential Linearized Augmented Plane Wave Package for Calculating Crystal Properties*, Karlheinz Schwarz, Techn. Univ. Wien, Vienna 1999, ISBN 3-9501031-0-4. Updated version of P. Blaha, K. Schwarz, P. Sorantin and S. B. Trickey; Comput. Phys. Commun., **59** 399 (1990).
[23] J. P. Perdew, K. Burke and M. Ernzerhof; Phys. Rev. Lett. **77**, 3865 (1996).
[24] A. R. Williams, V. L. Moruzzi, C. D. Jr. Gelatt, J. Kübler and K. Schwarz; J. Appl. Phys. **53**, 2019 (1982).
[25] K. Schwarz and P. Mohn; J. Phys. F: Met. Phys., **14**, L129 (1984).
[26] V. L. Moruzzi, P. M. Marcus, K. Schwartz, P. Mohn; Phys. Rev. B **34**, 1784 (1986).
[27] A. Neckel, K. Schwarz, R. Eibler and P. Rastl; Microchim. Acta, Suppl. **6**, 257 (1975).
[28] P. Ravindran, H. Fjellvåg, A. Kjekshus, P. Blaha, K. Schwarz and J. Luitz; J. Applied Phys. **91**, 291 (2002).
[29] R. P. Vasquez; Phys. Rev. B **54**, 14938 (1996).
[30] C. Zobel, M. Kriener, D. Bruns, J. Baier, M. Grüninger, T. Lorenz, P. Reutler and A. Revcolevschi; Phys. Rev. B **66**, 020402 (2002).